\documentclass[draft]{agujournal2019}
\usepackage{url} 
\usepackage{lineno}
\usepackage[inline]{trackchanges} 
\usepackage{soul}

\usepackage{amsmath,amssymb,amstext} 
\usepackage{enumerate}
\usepackage[authoryear]{natbib}
\usepackage[final]{pdfpages}

\draftfalse

\journalname{Geophysical Research Letters}

\definecolor{highlight}{rgb}{0, 0.2, .9}
\definecolor{darkblue}{rgb}{0.0, 0.0, 0.5}
\definecolor{lpur}{rgb}{0.8, 0.8, 0.95}
\definecolor{lyel}{rgb}{1.0, 1.0, 0.0}
\definecolor{gicol}{rgb}{1, 0.5, 0.0}
\definecolor{lgreen}{rgb}{0.5, 0.9, 0.7}

\newcommand{\difft}[1]{\frac{\mathrm{d}}{\mathrm{d} t} #1 } 
 
\newcommand{\pderiv}[2]{\frac{\partial #2}{\partial #1} }
\newcommand{\deriv}[2]{\frac{\mathrm{d}#2}{\mathrm{d} #1} }

\newcommand{\dz}{\ \hbox{d} z}

\newcommand{\TB}{T_B}

\newcommand{\FS}{F_S}
\newcommand{\SzH}{ S_{ z= H}}

\newcommand{\TA}{ T_{Air}}
\newcommand{\SB}{ S_B}
\newcommand{\SO}{ S_0}

\newcommand{\RG}{\mathcal{R}}

\newcommand{\FR}{\mathcal{F}}

\usepackage{appendix}

\begin{document}

%
%

\title{Salt-fingering in seasonally ice-covered lakes}

%
%




\authors{J. Olsthoorn \affil{1}, 
E. W. Tedford \affil{2}, 
G. A. Lawrence \affil{2}
}

\affiliation{1}{Department of Civil Engineering, Queen's University, Kingston, ON, Canada}
\affiliation{2}{Department of Civil Engineering, University of British Columbia, Vancouver, BC, Canada}





\correspondingauthor{Jason Olsthoorn}{Jason.Olsthoorn@queensu.ca}




\begin{keypoints}
\item Cryoconcentration produces gradients in solutes under the ice in freshwater lakes
\item Laboratory experiments indicate that salt-fingers, enabled by contrasting gradients in solutes and temperature, may be common in ice-covered lakes
\item Salt fingers may not obviously perturb the temperature stratification
\item Solutes excluded during ice-formation may become uniformly distributed with depth 
\end{keypoints}

%
%

%
%


\begin{abstract}
When ice forms on lakes, dissolved salts are rejected, which can lead to under-ice salt-finger formation. We performed a series of laboratory experiments to visualize these fingers. While we detected salt-fingers in our camera recordings, the signal of these fingers is nearly absent in the temperature record. We quantify the velocity of the salt plumes and measure the bottom salinity increase from these fingers. Further, we estimate that the salinity is often distributed evenly with depth. Comparing the salt fluxes in our experiments with a typical salt flux in lakes, we suggest that conditions are favorable for salt-fingering in most seasonally ice-covered lakes. 
\end{abstract}

\section*{Plain Language Summary}
When ice forms on the surface of lakes, dissolved salts are expelled from the ice into the liquid water below. If enough salt is rejected from the ice, the excess weight of the salt can lead to long `fingers` of salty fluid moving from the ice into the water below. We ran a series of experiments to investigate these `fingers', and conclude that this process likely occurs in most freshwater lakes that freeze annually. This process is important for the evolution of lakes and will change as fewer lakes freeze. 

%
%

\section{Introduction}\label{sec::Intro}

The past decade has seen a significant increase in studies on the physical processes occurring under lake ice. These processes had been previously understudied \citep{kirillin_physics_2012}, despite playing a key role in the yearly variability of important physical lake properties, such as temperature and oxygen \citep{yang_new_2021, cortes_mixing_2020}. While progress has been made, there remain many open questions concerning the physical, biogeochemical, and biological processes that are occurring under lake ice \citep{jansen_winter_2021}. One process in particular, cryoconcentration (or the rejection of solutes such as salts from freezing ice), may be an important process for both under-ice transport, and has the potential to suppress turnover in lakes \citep{pieters_effect_2009}. 

The diffusion rates of salt and heat in water are different. This difference can result in what are known as double-diffusive instabilities. Depending on how the salt and temperature are stratified, these instabilities will lead to two different types of flows. 
If the concentration of salt is unstably stratified (as is the case in cryoconcentration), and the temperature is stably stratified, double diffusive instabilities lead to the formation of long finger-like plumes called salt-fingers \citep{turner_buoyancy_2001}. These salt fingers have been observed in the ocean \citep{kunze_optical_1987} and play a key role in the vertical transport of heat in the Arctic \citep{rudels_double-diffusive_2009}. The other type of double-diffusive instability, diffusive convection, occurs when temperature is unstably stratified and the salt is stably stratified, such as in lake Kivu. For more information on double diffusive convection in lakes, see \citet{bouffard_convection_2019}.

The rejection of salt from ice in inland water (with salinities $<24$ppt) is complicated by a nonlinear equation of state. Unlike seawater, freshwater systems have a temperature of maximum density above their freezing temperature \citep{chen_thermodynamic_1986}. This nonlinear temperature dependence enables an \textit{inverse stratification} under ice, with colder-water above warmer water below. Despite this stable temperature stratification, sufficient cryoconcentration may lead to salt fingering, which is complicated by the nonlinear equation of state \citep{ozgokmen_asymmetric_1998}. Motivated by the work of \citet{bluteau_effects_2017}, \citet{olsthoorn_underice_2020} used numerical simulations to model these salt fingers and the resultant transport of salty water from the ice-water interface to the lake bottom. The present paper builds upon this previous work by performing a set of physical experiments to visualize the salt-exclusion-induced plumes.

However, salt-fingers are not the only physical process responsible for transporting salts under ice. \citet{mortimer_convection_1958} discussed how bottom heat flux and sediment respiration can generate to gravity currents that transport heat and solutes to the deepest portion of the lake. Furthermore, \citet{macintyre_sediment_2018} showed that even when sediment heat fluxes are low, sediment respiration can be enough to form these gravity currents. Calculating the true sediment respiration rate depends upon the chemistry of a lake, but, assuming a similar chemistry to \citet{mortimer_convection_1958}, \citet{deshpande_oxygen_2015} estimated that for their field site, sediment respiration and cryoconcentration had similar contributions to the increase in under-ice conductivity. The laboratory experiments presented in this paper do not include the effects of sediment respiration and will focus on cryoconcentration. 

Our paper has three objectives. First, we will visualize the salt-plumes and demonstrate their structure. We believe that we are the first to have done so in freshwater laboratory experiments. Second, we will highlight that salt-finger formation can occur even at very-low salinity values, and suggest that these plumes will occur in most ice-covered lakes. At low salinity, these plumes do not obviously perturb the temperature stratification, explaining why field measurements have often failed to identify their occurrence. Third, we highlight the role that these salt-plumes play in transporting dissolved solutes from the ice-water interface to the bottom of a lake.

\section{The Buoyancy Flux Ratio}\label{sec:parameter}

The density stratification under the ice is determined by temperature and salinity. Due to the nonlinear equation of state, the reverse temperature stratification under ice is stable. Cryoconcentration results in a flux of salt to the ice-water interface, thereby increasing the density near the water surface. This salt concentration is unstably stratified, meaning that in the absence of a temperature stratification, it would flow to the bottom of the domain. The relative strength of the temperature stratification to the salinity stratification then controls the type of under-ice flow that is observed.

The largest gradients of salinity and temperature are expected at the ice-water interface. At the interface, the salinity gradient is determined by the salt flux $F_s$ [ g kg$^{-1}$  m s$^{-1}$], and the temperature gradient is determined by the temperature flux ($F_T=-\kappa_0 \pderiv{z}{T}$ [$\ ^\circ$C~  m s$^{-1}$]). The relative strength of the salt stratification is then computed
\begin{gather}
    \RG = \frac{\beta F_S }{\alpha F_T},
\end{gather}
where $\alpha\equiv\alpha(T,T_{MD},S_0)\approx5.4-6.8 \times 10^{-5}$~[$\ ^\circ$C$^{-1}$] is the thermal expansion coefficient (ignoring variations in pressure), and $\beta\approx 8\times10^{-4}$~[kg g$^{-1}$] is the haline contraction coefficient, both of which were computed, at the freezing temperature, following \citet{chen_thermodynamic_1986}. Here, $\kappa_0\approx 1.3\times 10^{-7}$  m s$^{-1}$ is the molecular thermal diffusivity of water around 0 $\ ^\circ$C. Note that, if we define $\tau$ as the ratio of the thermal to saline diffusivities, the product $1/ \left(\tau \RG\right) = \alpha \pderiv{ z}{T} / \beta\pderiv{ z}{S}  = R_\rho$ is the density ratio typically defined in double-diffusive systems \citep{radko_double-diffusive_2013}. 
We will show that $\RG$ is the controlling parameter in this system.

%

The salt flux ($\FS$) at the ice-water interface is proportional to the rate of ice growth, and is calculated
\begin{gather}
    \FS = - r_S \SzH \deriv{t}{H},
\end{gather}
where $\SzH$ is the salinity at the ice-water interface, $H$ is the depth of the liquid water, and $r_s$ is the proportion of salt rejected from the ice. This salt-exclusion coefficient was $\approx 0.95$ in experiments with potassium chloride \citep{bluteau_effects_2017} and as high as 0.99 in field observations \citep{pieters_effect_2009}. We did not measure the salinity of the ice in the present experiments, and so we approximate $r_S = 1$. Next, as in \citet{olsthoorn_underice_2020}, we approximate that $\SzH\approx\SO$. Finally, we will show that ice grows approximately linearly for the experiments discussed in this paper. 
Combining these approximations, we model $F_S$ as a case dependent constant (see Table \ref{tab:CaseParameters}).

We briefly note that the experimental configuration described in this paper is different from the classic experimental setup of \citet{huppert_limiting_1973,ozgokmen_asymmetric_1998}. In those experiments, the total amount of dissolved solids was fixed, whereas in this case, cryoconcentration results in an increase in salt concentration over time. If this increase in salt were to continue indefinitely, we would expect to always see salt fingers. This is an important consideration when we discuss the laboratory experiments below. 

We will show that under-ice salt fingers transport salt away from the ice-water interface, leading to a increase in salt at the tank bottom ($F_B$). Will the transported salt pool at the bottom or will it be entrained into a salt stratification balanced by the stable temperature profile? We will discuss this question in \S\ref{sec:saltDistribution}, below.

\section{Laboratory Setup}\label{sec::Lab}

\begin{figure}
    \centering
    \includegraphics[width=1.0\textwidth]{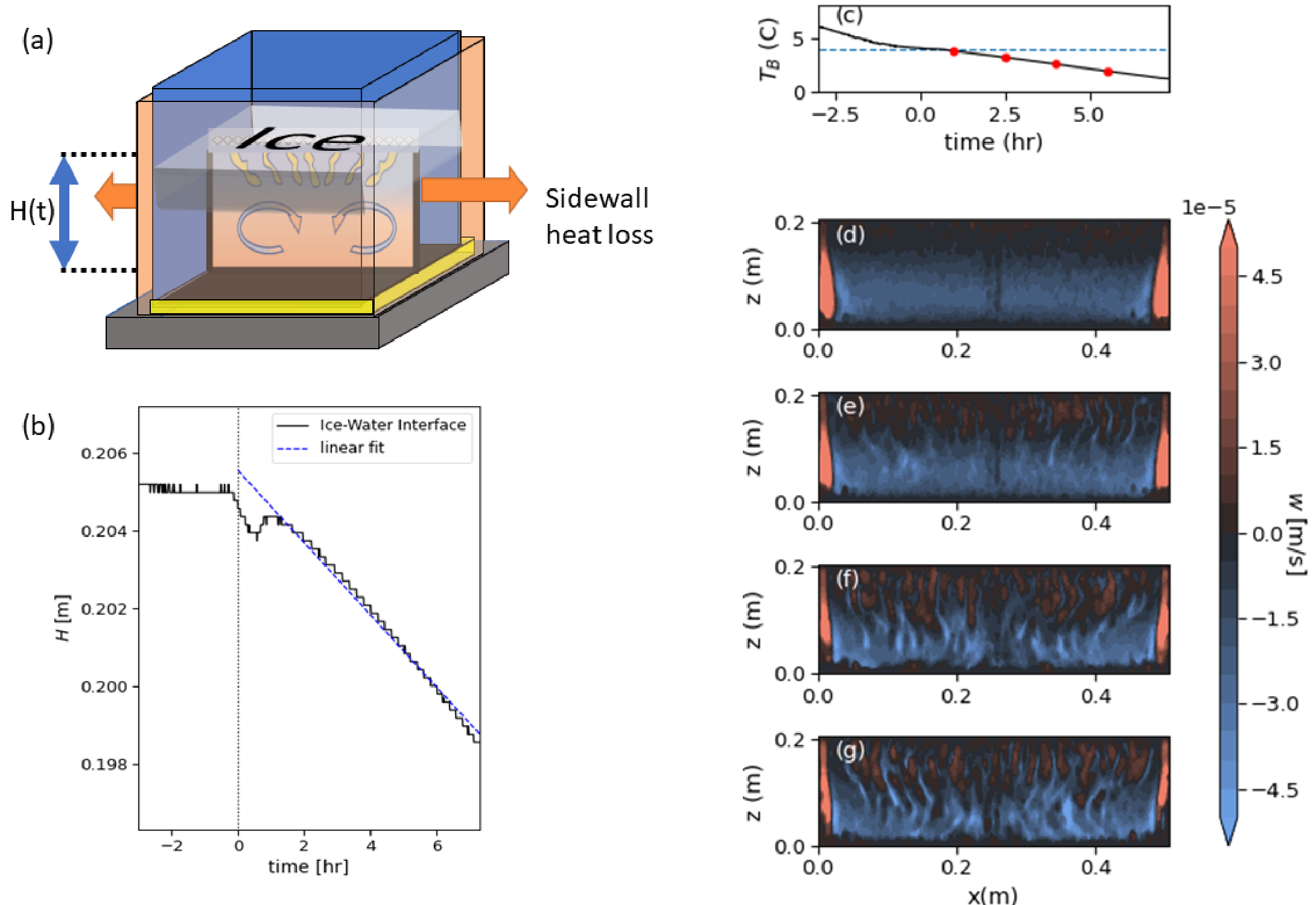}
    \caption{ (a) A schematic of the water tank that was placed into a freezer. The surface of the water was exposed to freezing atmospheric temperatures ($\TA<0\ ^\circ$C) leading to surface ice growth. Weak heat loss through the side-walls produced tank-scale recirculation cells below the growing ice. The rejected salts from the ice produced long finger-like plumes that propagate from the ice-water interface to the base of the tank. Note that the third acrylic side-wall was not included in the diagram. (b) Water depth as a function of time. Time $t=0$ is the time of ice formation. The initial dip in depth is a common artifact resulting from the change in water opacity when ice forms. Ignoring the defect, the water depth decreased approximately linearly with time.  (c) Temperature timeseries of the bottom water temperature ($T_B$) as a function of time. (d)-(g) Snapshots of the vertical velocity ($w$) at the times indicated by the red dots in panel (c). Data is presented for Case 4. }
    \label{fig:TankDiagram}
\end{figure}

We performed a series of laboratory experiments designed to visualize the salt-exclusion-induced convection in low-salt environments. These experiments were performed by placing a 50x50x50 cm$^3$ acrylic tank (Figure \ref{fig:TankDiagram}a) in a commercial walk-in freezer. The tank was partially filled with a sodium-chloride solution to a depth of $\approx20$ cm. The walls of the tank were triple-pane to thermally insulate the water from side-wall heat loss. The surface of the water was exposed to freezing atmospheric temperatures ($\TA<0\ ^\circ$C) leading to surface ice growth. While the dominant heat loss was through the surface, weak side-wall cooling did produce tank-scale recirculation cells between the centre and the sides of the tank.

To visualize the flow within the tank, we seeded the saline water with 20$\mu$m polyamid particles. We illuminated these particles with a vertical laser sheet (660nm) through the middle of the tank, which were then recorded at one frame per ten seconds. From these images, we computed the flow velocity with DigiFlow \citep{olsthoorn_three-dimensional_2017}. The speed of the slowest salt-fingers were comparable with the settling velocity of the particles ($\approx3\times 10^{-6}$  m s$^{-1}$). As we will discuss below, we compute the RMS velocity values to remove the error from both the settling velocity and the recirculation cells. 


We also recorded the temperature and salinity of the water. 
The water temperature was measured with an in-house built temperature chain, constructed from 10 BR14 NTC-type thermistors (Thermometrics Corp., CA), calibrated with an error of $\sim0.01 \ ^\circ$C.
These 10 thermistors were attached to a vertical rod at 2 cm spacing. One thermistor failed over the course of the experiments and was removed from the analysis. Further, these thermistors saturate at sufficiently low temperatures, at which point they were also omitted. The temperature chain was positioned in the center of the tank. 

A HOBO conductivity logger (Hoskin Scientific), positioned against the far wall of the tank, continuously measured the water conductivity. Further, we recorded reference conductivity values before and after each experiment with a handheld conductivity meter (Cole-Palmer Cond 3110). We converted between raw conductivity and specific conductance (C25, see table~\ref{tab:CaseParameters}), and subsequently salinity, following \citet{pawlowicz_calculating_2008}. A case-specific linear correction was also included to the Hobo record to correct for temperature-driven salinity drift prior to ice formation. The Hobo probe measured the salinity increase due to the salt excluded from the ice. 

In total, we ran 13 laboratory cases with initial salinity values of $\SO = $ 0 - 4  g kg$^{-1}$. Complementary experiments were performed with a freezer setpoint temperatures of $-10~\ ^\circ$C and $-15~\ ^\circ$C. Table  \ref{tab:CaseParameters} includes the individual case parameters. In each case, the camera images detected the ice formed on the water surface. We selected the time of ice formation as $t=0$. For freezer setpoint $-15~\ ^\circ$C, the ice formed nearly instantaneously across the water surface. For freezer setpoint $-10~\ ^\circ$C, often one side would freeze before the other (a time difference of up to 1hr) leading to an asymmetry in the salt exclusion rate. To account for this early asymmetric freezing, the data was omitted until the initial salt-plumes hit the bottom and the ice formation was more even. 

\begin{table}
	\begin{center}
		\begin{tabular}{c c c l l c c c c}
			Case & SetPoint  & $ H_0$ & C25  &  $S_0$   & $\Delta H$  & $F_S$   & $\mathcal{R}$ & Salt Fingers Obs.\\
			Case & [$\ ^\circ$C] &  [mm] & [mS cm$^{-1}$] &   [ g kg$^{-1}$]  &  [mm] &  [$10^{-9}$  g kg$^{-1}$  m s$^{-1}$]  & [-] & \\
			\hline
			0 & -15 & 223 & 0.002 & 0.001  & 7.0 & 0.2 & -- & No\\
			1 & -15 & 215 & 0.029 & 0.013 & 7.8 & 3.3 & 0.03-0.04 & No\\
			2 & -15 & 212 & 0.047& 0.022 & 8.0 & 6.0 & 0.04-0.08 & Yes\\
			3 & -15 & 187 & 0.081& 0.038 & 7.4 & 10.3 & 0.05-0.12 & Yes\\
			4 & -15 & 205 & 0.18& 0.086 & 6.6 & 22.0 & 0.08-0.18 & Yes\\
			5 & -15 & 213 & 0.72 & 0.35 & 7.2 & 99.4 & 0.24-1.20 & Yes\\
			6 & -15 & 220 & 2.4 & 1.2 & 6.7 & 270 & 0.45-3.07 & Yes\\
			7 & -15 & 226 & 6.5 & 3.5 & 4.6 & 660 & 0.70-6.91 & Yes\\
			8 & -10 & 193 & 0.078 & 0.037 & 4.9 & 6.1 & 0.02-0.04 & Yes\\
			9 & -10 & 215 & 0.17 & 0.082 & 5.4 & 12.9 & 0.06-0.13 & Yes\\
			10 & -10 & 216 & 0.72& 0.35 & 4.0 & 51.2 & 0.14-0.44 & Yes\\
			11 & -10 & 210 & 2.3 & 1.2 & 2.7 & 151 & 0.35-0.73 & Yes\\
			12 & -10 & 224 & 6.6 & 3.5 & 2.9 & 460 & 1.06-3.00 & Yes
		\end{tabular}
	\end{center}
	\caption{Table of the basic parameters associated with each experiment. The initial water depth and initial mean water salinity was $H_0$ and $S_0$, respectively. The specific conductivity (C25) was standardized at 25$\ ^\circ$C. The net decrease in water depth over each experiment is denoted $\Delta H$.}
	\label{tab:CaseParameters}
\end{table}


\section{Results} \label{sec::Results}

\subsection{Ice Growth}

We measured the mean water depth $H$ over time from the camera images (Figure \ref{fig:TankDiagram}b). Before ice formation ($t<0$), $H$ is constant; as ice grows ($t>0$), $H$ slowly decreased.  Due to the change in opacity of the ice as it forms, the detection algorithm often struggled to detect the initial growth of ice (see the blip around $t=0.5$hr in Figure \ref{fig:TankDiagram}b). This was worse at high salinity values. The ice growth was approximately linear in time for all cases, as argued by \citet{ashton_thin_1989} for thin ice, and was primarily a function of the freezer setpoint temperature. 
Typical final ice thicknesses were around $0.5$cm.

\subsection{Time Evolution}

We discuss the time evolution of these experiments and the generation of salt fingers. As a representative case, we will initially focus on Case 4, before comparing this case with the others. 

 After ice formation, the reverse temperature stratification continued to cool in all cases. Figure \ref{fig:TankDiagram}c is a plot of the bottom water temperature ($\TB$) over time for Case 4. Initially ($t\approx 0$) the water preferentially cooled near the surface such that the bottom water temperature $\TB$ was approximately constant. As the system continued to cool, $\TB$ decreased.


Figure \ref{fig:TankDiagram}d-g contains snapshots of the vertical velocity $w$ at the times indicated by the red dots in Figure \ref{fig:TankDiagram}c. 
Prior to the formation of salt-fingers, side-wall heat losses generated tank-scale recirculation cells (see Figure \ref{fig:TankDiagram}d). 
In particular, the heat loss at the side-walls induced a negatively buoyant jet (red), which then induced a counter-flow of the bulk interior fluid moving down (blue). Note that the thermistor chain in the domain centre ($x\approx 0.25$cm) interferes with the PIV and the affected region is removed in our subsequent analysis. 


By $t=2.5$ hr, we observe the formation of finger-like plumes (Figure \ref{fig:TankDiagram}e), which grow at the surface and propagate down towards the tank bottom (Figure \ref{fig:TankDiagram}f-g). These plumes induce their own counter-flow between the fingers. The recirculation cells enhance the downward propagation of the plumes.
Importantly, these plumes transport salt from the ice-water interface to the tank bottom. In a lake, these plumes would also transport oxygen, nutrients, contaminants and other dissolved constituents.

\subsection{Heat Transport}
\begin{figure}
\centering
  \includegraphics[width=1.0\textwidth]{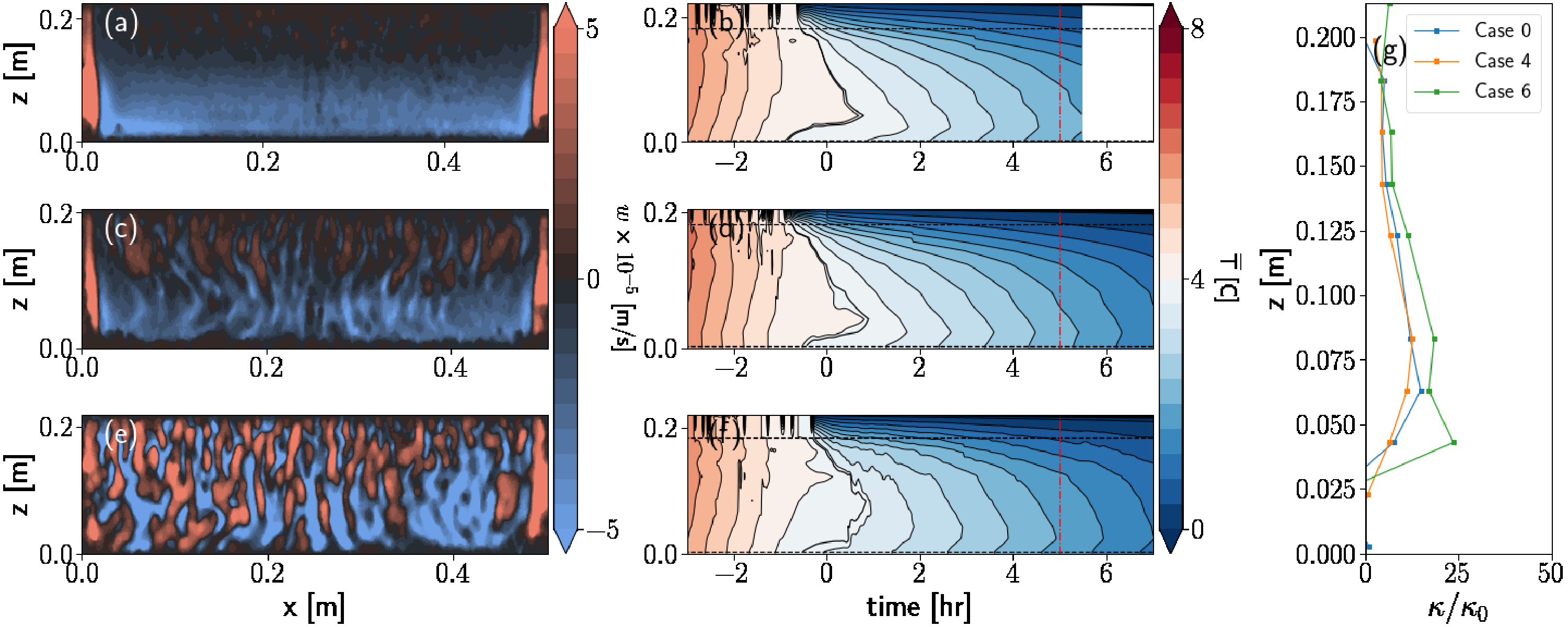}
\caption{Plot of the vertical velocity (a),(c),(e) and temperature stratification (b),(d),(f) for Cases 0, 4, and 6. As the water salinity increases, the intensity of the down-welling plumes increases. The vertical velocity snapshots were output from time $t$=$5$ hrs, indicated by the vertical red line on panels (b),(d),(f). The scaled turbulent temperature diffusivity has also been included in panel (g).}
    \label{fig:VelT_Comp}
\end{figure}

The evolution of the observed plumes depend upon the salt-exclusion rate. Figure \ref{fig:VelT_Comp} contains contour plots of (a,c,e) the vertical velocity and (b,d,f) the temperature stratification for Cases 0, 4, and 6. These cases covered a range of water salinity from 0.0 -- 1.3  g kg$^{-1}$, and highlight the difference between distilled, low-salinity, and medium-salinity environments. The horizontal dashed lines in the temperature plots (Figure \ref{fig:VelT_Comp}b,d,f) denote the position of the top and bottom thermistor, with temperature above the top thermistor interpolated with an error-function. Before ice formation ($t<0$), the water temperature was well mixed due to surface driven convection. After ice formation, we observe a reverse temperature stratification that continued to cool during the remainder of the experiment. 

We ran an experiment with distilled water (Case 0, $S_0 \approx 0$  g kg$^{-1}$) 
to underscore that the observed plumes result from salt exclusion.  As expected, no plumes were observed in the absence of salt (Figure \ref{fig:VelT_Comp}a) and the temperature stratification cooled smoothly. 

For low-salinity environments (Case 4, $S_0 = 0.09$  g kg$^{-1}$, as with Figure \ref{fig:TankDiagram}), salt fingers form (Figure \ref{fig:VelT_Comp}c) but they are sufficiently weak such that the temperature evolution is nearly unchanged (Figure \ref{fig:VelT_Comp}d). 
Salt exclusion produced salt fingers in very low salinity environments; salt fingers are not always identifiable in the temperature record. 

As the salinity increased (Case 6, $S_0 = 1.2$  g kg$^{-1}$), salt exclusion resulted in more energetic plumes (Figure \ref{fig:VelT_Comp}e). These more energetic plumes generated convective motions that perturbed the temperature stratification (see Figure \ref{fig:VelT_Comp}f around $t=1$ hr). As the initial bulk salinity $S_0$ increased, the speed of the plumes increased and induced larger perturbations to the temperature field. The plumes also form earlier (See Supplementary Information A).

Both the sidewall convection and the saline plumes stirred the water, which enhanced the vertical transport of heat. We estimated the magnitude of this effect through an average turbulent thermal diffusivity $\overline \kappa_*$ (See Supplementary Information B for more details). 
Figure \ref{fig:VelT_Comp}g is a plot of $\overline \kappa_*$ for each of the three plotted cases. By investigating Case 0 (distilled water), we can distinguish the contribution of the heat transport from the sidewall convection and the saline plumes. As shown in Figure \ref{fig:VelT_Comp}g,  
 the sidewall convection increased $\overline \kappa_*$ by a factor of 10 from its molecular value $\kappa_0=1.3 \times 10^{-7}$ m$^2$ s$^{-1}$. This sidewall convection dominates the heat transport when the water salinity is very low. For example, Case 0 and 4 have approximately the same thermal diffusivity. 
 As the speed of the saline plumes increased, $\overline \kappa_*$ further increased by a factor of 2 above the pure sidewall-convection value. Thus, at moderate salinity, the saline plumes do increase the heat transport in the system but this is likely to be subdominant to other physical processes occurring in a natural system. This does not mean that the \textit{salt} transport is negligible. Next, we will show that the plumes increase the salt transport even at low water salinity.

\subsection{Salinity Increase}
\begin{figure}
    \centering
    \includegraphics[width=1.1\textwidth]{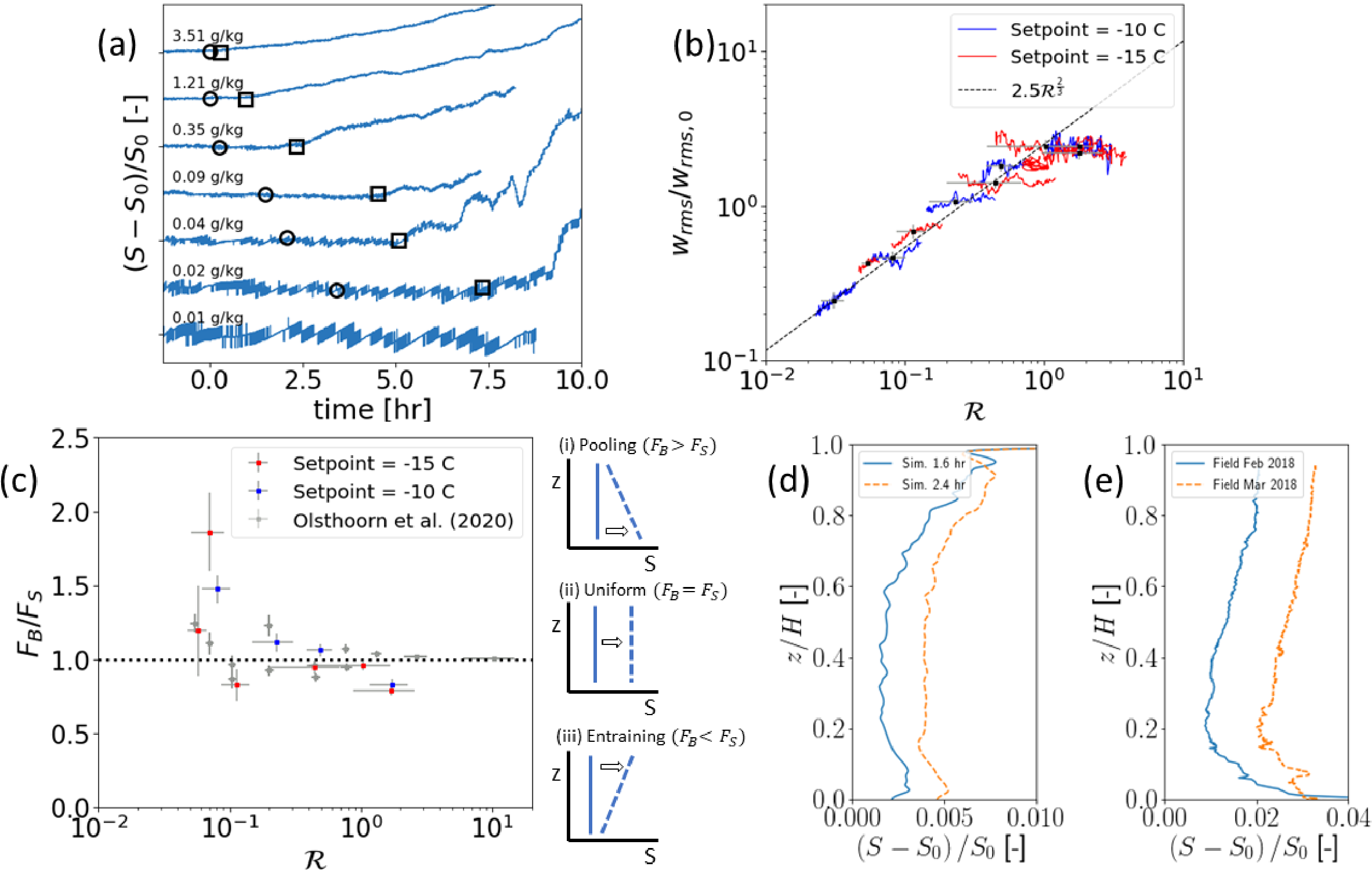}
    \caption{ (a) Salinity perturbation as a function of time for Cases 1-7. The black squares indicate the time when the salinity perturbation starts to increase ($t_M$). Circles are denoted as the estimated time of finger formation ($t_f$). (b) Vertical velocity RMS as a function of $\RG$. For low $\RG\lesssim 1$, the $w_{rms}$ scales as $\RG^\frac 23$. For $\RG \gtrsim1$, $w_{rms}\approx 2 w_{rms,0}$. (c) The flux ratio $\FR$ plotted as a function of the buoyancy ratio $\RG$. The horizontal dashed line is given at $\FR=1$ . The estimated transition from (i) pooling to (ii) uniform distribution is at $\RG=0.5$. (d) Two simulated salinity profiles from \citet{olsthoorn_underice_2020} (Case 5, $S_0=1$ g kg$^{-1}$, $H=0.1$m). (e) Two field salinity measurements from BML ($S_0 = 1.75$ g kg$^{-1}$, $H=9.7$m). }
    \label{fig:ResultsCollection}
\end{figure}

The salt-plumes transport salt from the ice-water interface to the water below. Eventually, the salt-fingers reach the tank bottom, where the salinity ($\SB$) was continuously measured. Figure \ref{fig:ResultsCollection}a is a plot of the salinity perturbation $S^\prime = \left( \SB - \SO\right) / \SO$ for Cases 1-7 as a function of time. The data have been offset to reveal the temporal variation of the seven cases. Notice that after some time, the bottom water salinity starts to increase.
We identify this time $t_M$, using a piecewise linear curve, which is indicated by the black squares in Figure \ref{fig:ResultsCollection}a. This measurement time $t_M$ identifies when the rejected salt first reaches the tank bottom.

The initial time of salinity increase 
 \begin{gather}
     t_M = t_f + \Delta t_p
 \end{gather} 
 is given by the finger formation time ($t_f$) plus the propagation time ($\Delta t_p$) for the salt fingers to reach the bottom of the tank.  
 We estimate the formation time $t_f$ through back-propagation of the first fingers, and denote the $t_f$ estimate with circles in Figure \ref{fig:ResultsCollection}a. 
 We find that the plumes form earlier at higher salinity values, consistent with the prediction that the time required for the salt-plumes to form depends on the salt-exclusion rate (see Supplementary Information A). At the highest salinity values, the fingers formed nearly instantaneously with ice formation; $t_f\approx 0$. 
The propagation time ($\Delta t_p = t_M - t_f$) for the plumes to reach the tank bottom similarly decreased with salinity. That is, the plumes travelled more quickly with higher salinity (see  \S \ref{subsec::RMS}). 

It is worth highlighting that for $S_0 = 0.01$  g kg$^{-1}$, no plumes were detected before the experiment ended. Similarly, the salinity did not increase above its initial value. 
Nevertheless, we observed that salt-fingers at salinity values as low as 0.02  g kg$^{-1}$, at the low end of the limnological range of salinity (see \S\ref{sec::Lakes} for more details).
Based upon this observation, we suggest that salt-exclusion-induced salt fingers are a significantly more ubiquitous under-ice process than has been previously recognized.

\subsection{RMS Velocity} \label{subsec::RMS}

 The saline plumes propagate faster at higher $S_0$. As noted in \S \ref{sec:parameter}, the stable temperature stratification also controls the decent of these plumes. While the temperature stratification does vary with depth (see Figure \ref{fig:VelT_Comp}b,d,f), the instability growth rate is controlled by the temperature stratification and salt flux, at the ice water interface. The controlling parameter of the system is $\RG=\frac{\beta F_S }{\alpha F_T}$. 

We compute average plume speed as the root-mean-squared vertical velocity $w_{rms}$ over the domain. We applied a high-pass spatial filter to remove the effect of the recirculation cells. In addition, we removed the velocity data within 5cm of the side-wall due to the boundary layers, along with the data affected by the thermistor chain.  Further, we filter out data for $t<0.5$ hr to avoid any remnant pre-ice convection. We decided to use this simple filter as more sophisticated algorithms, such as Dynamic Mode Decomposition, produced similar results. Figure \ref{fig:ResultsCollection}b is a plot of $w_{rms}$ as a function of $\RG$. The median values for each case are plotted as black squares, and the error bars are the standard deviation along each axis. In this plot, we scaled $w_{rms}$ by the appropriate reference velocity $w_{rms,0} = \frac{\kappa_T}{d}$, where length scale $d$ is the finger width \citep{radko_double-diffusive_2013}. We determined, from an integral measure of the velocity spectra, that $d\approx 7.5$ mm in all cases (See Figure \ref{fig:TankDiagram}g). This value is consistent with the field measurements of salt fingers in the Western Atlantic \citep{kunze_optical_1987}. We estimate $w_{rms,0} \approx 1.8 \times 10^{-5}$  m s$^{-1}$.

We observe that $w_{rms}$ increases with $\RG$. 
For large values of $\RG\gtrsim 1 $, $w_{rms}$ appears to saturate with a value of $w_{rms}/w_{rms,0} \approx 2$. We associate this saturation with the expected regime change around $\RG\approx 1$ from double-diffusive fingers to salt-driven convection \citep{olsthoorn_underice_2020}. \citet{olsthoorn_underice_2020} similarly observed a saturation in the relative heat flux and associated it with the same regime transition.

For $\RG<1$, we find that the $w_{rms}/w_{rms,0}$ scales with $\sim \RG^{\frac 23}$. We leave theoretical explanation for this scaling for future work. However, this scaling indicates that $\RG$ is indeed the correct controlling parameter and that the plume velocities increase sub-linearly with increasing salt flux.

\subsection{Salt distribution} \label{sec:saltDistribution}

Salt-fingers transport salt through the background thermal stratification. As the plumes descend, the transported salt may 
\begin{enumerate}[(i)]
\item pool at the bottom creating a stable bottom saline layer,
\item be uniformly distributed with depth, or 
\item be continuously entrained into an unstable salt stratification.
\end{enumerate}
The stable temperature stratification can support an unstable salinity stratification while remaining hydrostatically stable. 
However, option (iii) is unlikely to occur as the salt stratification would rapidly overcome the stabilizing effect of the temperature stratification. We can determine how salt is being transported by comparing the amount of salt excluded from the ice with the measured amount of salt that reaches the tank bottom. 
Assuming that the salt is completely rejected from the ice,
 salt is conserved in the water such that 
\begin{gather}
    \difft{\int_0^H S \ \dz} =0  \implies F_s = \underbrace{H\difft{S_{B}}}_{F_{B}}  + \underbrace{\int \pderiv{t}{} \left( S - S_{B}\right)\ \dz}_{Excess/Deficit}. \label{eqn:saltbudget}
\end{gather}
Here, $F_B$ is the salt flux estimate assuming the water were uniformly mixed to $S_B$.
That is, if the plumes deposit salt uniformly with depth, the salinity flux ratio
\begin{gather}
    \FR = \frac{F_{B}}{F_S} = \frac{H \difft{S_{B}}}{F_S} = 1. \label{eqn:SFR}
\end{gather}
The final term in \eqref{eqn:saltbudget} represents the salinity deficit or excess associated with either (i) salt pooling  at the bottom of the tank ($\FR>1$), or (iii) entraining into the stratification ($\FR<1$). 

We want to highlight that $\FR=1$ does not imply that the salt stratification is uniform. Rather, a uniform salt deposition will keep the shape of the salinity profile steady, while increasing in time. As we will discuss in \S \ref{sec::Lakes}, there are field observations of salinity profiles that maintain their shape over time, while becoming increasingly saline.

Figure \ref{fig:ResultsCollection}c is a plot of $\FR$ as a function of $\RG$. The error bars are the standard deviation of the scaled salinity error from linear growth, and the standard deviation of $\RG$ over the experiment. We note that both axes contain the salt-exclusion rate $F_S$, and, in that sense, are not orthogonal quantities. We include the data from the two-dimensional numerical simulations from \citet{olsthoorn_underice_2020}. For the numerical data, the mean salinity in the bottom 5\% of the domain was used as a numerical proxy for $S_{B}$. We observe that to a reasonable approximation, $\FR\approx 1$ for most cases, with some scatter. For low $\RG$, the data are biased towards the pooling regime where $\FR\gtrsim 1$. Due to the large spread in the data, we do not assign a precise scaling law for $\FR$ vs $\RG$. Instead, we simply estimate that the transition between (i) pooling ($\FR\gtrsim 1$) and (ii) uniform deposition ($\FR\approx1$) occurs at $\RG \approx 0.5$.

\section{Implications for Lakes}\label{sec::Lakes}

We have shown that salt fingers can form in water with very low salinity. Although others have shown that cryoconcentration (salt-exclusion) can lead to an increase in salinity under ice \citep{pieters_effect_2009}, we are the first to visualize the mechanism for this salt transport. We are also the first to show that salt-exclusion leads to salt finger formation in water with a salinity as low as 0.02  g kg$^{-1}$. We will argue that salt-exclusion-induced salt fingers are common in natural lake settings.

As discussed in \S\ref{sec:parameter}, the relevant parameter for salt-finger formation is $\RG$. The thermal stratification in the laboratory experiments is stronger than typically found in lakes under ice. Survey data from \citet{malm_temperature_1997} measured under-ice temperature gradients of $\partial T/\partial z \approx 1-5\ ^\circ $C~m$^{-1}$, and \citet{cortes_mixing_2020} report a values as high as $2\ ^\circ $C~m$^{-1}$. The present experiments typically have $\partial T/\partial z \approx 9-60\ ^\circ $C~m$^{-1}$, suggesting that lake values of $\RG$ would be even larger than these experiments with similar salt flux $\FS$. That is, we would expect to find salt-fingers in lakes with lower values of $\FS$ than in the laboratory experiments. We will first discuss estimates of the salt-exclusion rates found in lakes, before returning to discuss estimates of $\RG$.

\citet{lerman_physics_1995} estimated the global salt concentration for lakes (excluding saline lakes with salinity $>3$  g~kg$^{-1}$) to be 0.2  g~kg$^{-1}$. 
As an example of a natural ice growth rate, the freezing rate from the St. Lawrence field data of \citet{ashton_thin_1989} is $\approx 0.2$  mm hr$^{-1}$ over 10 days. Similarly, we estimate a freezing rate of $\approx 0.3-0.4$  mm hr$^{-1}$ over the first two months of ice growth from the data of \citet{yang_modelling_2013}. Taking $0.2$  mm hr$^{-1}$ for thin ice growth over a day, we predict a salt flux of $ 1 \times 10^{-8} $  g kg$^{-1}$  m s$^{-1}$. We have observed salt-exclusion-induced salt-fingers for salt fluxes at a third this value within several hours. \citet{malm_temperature_1997} suggested that essentially no black-ice formed in Lake Vendyurskoe between December through April, estimating $\RG\approx 0$ and no salt finger generation. However, this would imply that all of the 0.5m of ice formed between November and December, giving an estimated $F_S \approx 5\times 10^{-9}$  g kg$^{-1}$  m s$^{-1}$ for this extremely fresh lake ($S_0 \approx 0.02$  g kg$^{-1}$, assuming two months of ice growth). 

While ice grows most quickly early in the season, sometimes growing almost exclusively in the first few months \citep{malm_temperature_1997, yang_mixing_2020,malm_temperature_1997}, even seasonally averaged salt-exclusion rates (an underestimate of the true rate) are typically greater than the values found in our experiments, suggesting that salt-fingers will form in most ice-covered lakes. For example, Boreal lakes have a mean ice cover of 0.5m over the winter \citep{kirillin_physics_2012}. Assuming the ice grows continuously over the winter (e.g. six-months), this slowest rate at which the ice could form would predict $F_s = 6.4\times 10^{-9}$  g kg$^{-1}$  m s$^{-1}$ ($S_0 = 0.2$ g kg$^{-1}$ by 0.5 m / 6 months). This bulk estimate is roughly consistent with the estimated mid-winter ice growth for BML \citep{chang_heat_2020} and other lakes \citep{oveisy_one-dimensional_2014, yang_highfrequency_2017}. We have observed salt-fingers (Case 2) when the salt flux was still lower than this minimum estimate. As the ice growth in lakes is significantly greater in the beginning of winter, we suggest that salt finger will form in most Boreal lakes for some portion of the season.

It is challenging to evaluate $\RG$ from the literature as it requires a simultaneous estimate of the ice growth rate, bulk water salinity ( g kg$^{-1}$), and under-ice temperature gradient. Nevertheless, the work of \citet{bengtsson_field_1996} provides an estimate of the under ice temperature gradient to be 1.6 $\ ^\circ$C m$^{-1}$. If we use the lower-bound estimates for the freezing rate of Boreal lakes, we estimate $\RG \approx 0.4$.  Further, \citet{malm_temperature_1997} measured under-ice temperature gradients in two other lakes ranging from 0.7-13 $\ ^\circ$C m$^{-1}$, which, assuming a similar minimal salt flux to the above, provides a range of $\RG\approx 0.04-0.8$. 
 While these numbers are highly variable, these estimates are consistent with the conclusion that salt-fingers will form beneath lake ice.



These salt-fingers move slowly, with speeds of $O(10^{-5}$  m s$^{-1}$)(see Figure \ref{fig:ResultsCollection}b), and they have a width of $\approx 1$cm (see Figure \ref{fig:VelT_Comp}c). As discussed in \S \ref{subsec::RMS}, these values are consistent with the field measurements of \citet{kunze_optical_1987}. Salt-exclusion from ice growth is responsible for finger formation. However, even when ice stops growing, these fingers may continue to descend. Over a day, the plumes will descend a distance of $O(1 $m), and possibly up to $O(100 $m) over the winter. In the absence of field measurements, it is unknown how other processes will change the salt finger propagation. 


The laboratory experiments described in this paper do omit several physical processes found in environmental lake systems. Namely, we do not include the effect of sediment respiration \citep{mortimer_convection_1958}, which can increase under-ice salinity \citep{macintyre_sediment_2018}. Sediment respiration increases the salinity at the lake bottom. Conversely, salt fingers form at the ice-water interface and propagate down to the lake bottom. 
\citet{malm_temperature_1997} estimated the average salt flux from sediment respiration in Lake Vendyurskoe to be $\approx 0.5 \times 10^{-9}$  g kg$^{-1}$  m s$^{-1}$. However, they noted that the salinity increase due to sediment respiration over six months could be accounted for by salt release from the ice.
It is not yet known how these two processes will interact. 

Another important process that we have not included is radiatively driven convection \citep{ulloa_mechanical_2018}. This process is primarily important at the end of winter \citep{kirillin_physics_2012}, and typically after most of the ice has formed. Radiatively driven convection often results in a mixed surface layer under the ice, with convective velocities that are orders of magnitude larger than that of the salt fingers. 
Similarly, we do not include the effects of under-ice internal waves \citep{cortes_mixing_2020}. Finally, the freezer temperature was fixed (modulo periodic oscillations from the freezer's compressor), while daily atmospheric temperatures and incoming radiation can vary significantly, particularly in late winter. 
Nevertheless, we suggest these additional parameters primarily affect the surface salt-flux $F_s$, and hence $\RG$, modifying the evolution of the salt fingers, but not necessarily eliminating them.


The salt fingers transport dissolved constituents faster than molecular diffusion. As the fingers propagate, some of the constituents mix into the water column, potentially generating a stratification. We defined the salt flux ratio, $\FR$, in \eqref{eqn:SFR}, to quantify how the salt stratification evolves in time. Measuring $\FR\approx1$ indicates that the surface salt flux is distributed evenly with depth. This is supported by the 2D numerical results of \citet{olsthoorn_underice_2020}, which show that the structure of the salt stratification remained nearly fixed, while the salinity increased over time (Figure \ref{fig:ResultsCollection}d).
A shift in the salinity stratification is also observed in field measurements. The mean salinity of Base Mine Lake ($S_0\approx 1.75$  g kg$^{-1}$), a constructed lake in central Alberta, Canada \citep{tedford_temporal_2019}, increased by approximately 1\% between Feb 6th, 2018, and March 7th, 2018 (see Figure \ref{fig:ResultsCollection}e).  
 We estimate $\approx6$cm of ice growth over that period (see chapter 5 of \citet{chang_heat_2020}) accounting for $\approx60$\% of the salinity increase. 
 The remaining salinity increase can be accounted for by pore water expression \citep{dompierre_chemical_2017} and sediment respiration \citep{mortimer_convection_1958}. We find that the shape of the salinity profiles are consistent after a month, while the average value has increased under the ice, suggesting $\FR\approx 1$. 
 
 Not all lakes exhibit such a uniform shift in under-ice salinity. \citet{malm_temperature_1997a} contains temperature and salinity profiles for three ice-covered lakes over the winter period. Many of these profiles have a measurable increase in salinity near the ice-water interface (see their Figure 10). However, the salinity profiles exhibit a ``pooling" type of salinity increase, with salinity preferentially increasing near the bottom. Without an estimate of the ice growth, we do not have an estimate of $\RG$ and we do not know if this evolution is a result of salt-fingers, sediment respiration, or some combination of the two. Similarly, it is not yet know how the salt-fingers would interact with a pre-existing stable salt stratification. It is possible that the descending plumes would be arrested by, and enhance, the existing stratification. Further studies are required to characterise this effect. 




Certainly, cryoconcentration is expected to play a larger role for inland waters with higher salinity, such as those impacted by mining \citep[e.g. tailings ponds, see][]{Lawrence2016}. However, even in low-salinity lakes, salt exclusion can produce salt fingers that are able to penetrate the temperature stratification and transport water from the ice-water interface down to the lake bottom. Other under-ice mixing processes, such as radiatively driven convection \citep{kirillin_physics_2012,ulloa_mechanical_2018}, are localized near the ice-water interface and necessarily rely on mixing of the temperature stratification. As highlighted by \citet{yang_mixing_2020}, weak salinity gradients at depth can balance the buoyancy effects of the temperature stratification and thus, even `small' salt fluxes can effect the stratification at lake bottoms. Our manuscript highlights that salt-exclusion may play a role transporting salt, and other dissolved constituents, to otherwise inaccessible regions.

\section{Conclusion}\label{sec::Conclusions}

 We have shown that salt-exclusion enhances salt transport between the ice-water interface and lake bottoms. We argue that salt-fingering occurs in most seasonally ice-covered lakes. As lakes continue to evolve by both warming \citep{oreilly_rapid_2015}, losing ice cover \citep{magnuson_historical_2000}, and becoming increasingly saline \citep{dugan_salting_2017}, more field monitoring is required to understand how salt-exclusion will affect salt transport in lakes in the future.  


\section*{Acknowledgement}
 Funding for this research was provided by Natural Sciences and Engineering Research Council of Canada (NSERC) and Syncrude Canada Ltd. Additionally, Jason Olsthoorn was supported by funding from the Killam Trusts. The Authors would like to thank the two anonymous Referees for their helpful input that improved this paper. 
 
 \subsection*{Open Research}
The processed data used for the analysis in the study are publicly available. The data is hosted on Borealis though Queen's University Dataverse 

(DOI:~https://doi.org/10.5683/SP3/8KQWHS).

\bibliography{bib}

\includepdf[pages=-]{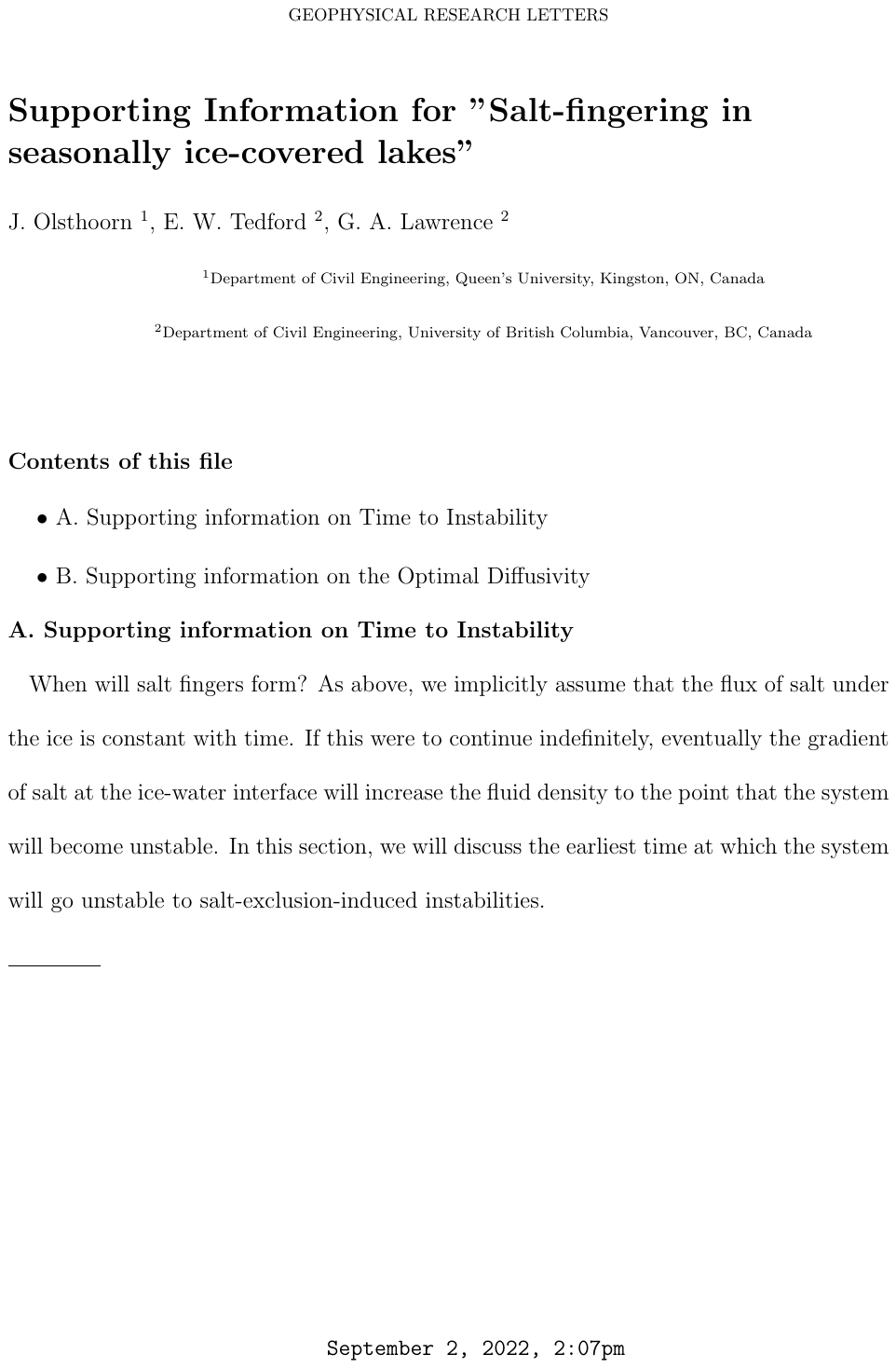}

\end{document}